\documentclass[journal]{IEEEtran}
\usepackage{cite}

\usepackage{graphicx}
\ifCLASSINFOpdf
\else
\fi
\usepackage[cmex10]{amsmath}
\usepackage{array}

\usepackage{mdwmath}
\usepackage{mdwtab}

\usepackage[font = small]{caption}
\begin{document}
%
\title{Dispersion Compensation using High-Positive Dispersive Optical Fibers}
%
%
%
\author{Mohammad~Hadi,
        Farokh~Marvasti,
        Mohammad~Reza~Pakravan

                
\thanks{M. Hadi is a PhD Candidate at Electrical Engineering Department, Sharif University of Technology, E-mil: mhadi@ee.sharif.edu}
\thanks{F. Marvasti is director of ACRI and fauclty member of Electrical Engineering Department, Sharif University of Technology, E-mil: marvasti@sharif.edu}
\thanks{M. R. Pakravan is faculty member of Electrical Engineering Department, Sharif University of Technology, E-mil: pakravan@sharif.edu}}

\maketitle

\begin{abstract}
The common and traditional method for dispersion compensation in optical domain is concatenating the transmit optical fiber by a compensating optical fiber having high-negative dispersion coefficient. In this paper, we take an opposite direction and show how an optical fiber with high-positive dispersion coefficient can also be used for dispersion compensation.  Our optical dispersion compensating structure is the optical implementation of an iterative algorithm in signal processing. The proposed dispersion compensating system is constructed by cascading a number of compensating sub-systems and its compensation capability is improved by increasing the number of embedded sub-systems. We also show that the compensation capability is a trade-off between transmission length and bandwidth.  We use simulation results to validate the performance of the introduced dispersion compensating module. Photonic crystal fibers with high-positive dispersion coefficient can be used for constructing the proposed optical dispersion compensating module.
\end{abstract}


%
\IEEEpeerreviewmaketitle

\section{Introduction}\label{Sec_I}
%
%
%
%
\IEEEPARstart{S}{ince} the revolution of fiber optics communication in decade $90$, millions of kilometers of optical fibers have been laid all over the world to convey the ever growing data streams primarily driven by the exponential increase in communicated video and data traffic. This exponential growth in data traffic is met through increasing per-channel bit rate or number of accommodated sub-channels using techniques such as Wavelength Division Multiplexing (WDM) or Optical Orthogonal Frequency Division Multiplexing (O-OFDM) \cite{cvijetic2012ofdm, zhang2013survey}. Increasing per-channel bit rate needs transmission of narrower optical pulses which multiplies signal degradation due to high amount of accumulated chromatic dispersion during pulse propagation through the optical fiber. This high amount of dispersion may result in Inter-Symbol Interference (ISI), information loss and increased Bit Error Rate (BER) value. on the other hand, increasing the number of accommodated sub-channels results in reduced space between adjacent sub-channels, more transmitted optical power and more sensitivity to destructive fiber nonlinear effects. Consequently, both of increasing per-channel bit rate and number of accommodated sub-channels lead to excessive sensitivity to certain fiber impairments that degrade clean transmission of the optical signal \cite{thyagarajan2007modeling}. Practically, the increased optical power along with tighter sub-channel spacing and longer transmission distance are translated to a trade-off between nonlinear propagation effects and chromatic dispersion in a fiber optics communication system. Today's, advanced optical fibers are designed such that they exhibit finite dispersion in the transmission band. This finite amount of dispersion reduces the growth of non-linear effects such as  Four-Wave Mixing (FWM) and Cross-Phase Modulation (XPM) which are particularly deleterious in DWM and O-OFDM communication systems \cite{agrawal2007nonlinear, itu2016}. In order to resolve the ISI and BER forced by the mentioned finite amount of dispersion, proper dispersion compensating techniques have been proposed to compensate for the accumulated dispersion in the propagated pulse through the optical fiber. Compensation can be achieved in the optical domain by the use of different dispersion compensation devices such as Dispersion Compensating Gratings (DCG),  Dispersion Compensating Fibers (DCF), Dispersion Compensating Arrayed Waveguides (DCAW), etc \cite{thyagarajan2007modeling, ramachandran2007fiber, gerome2004design}. Electronic Dispersion Compensators (EDC) are also proposed to compensate for  the accumulated dispersion in the electrical domain \cite{bulow2008electronic, schmidt2008experimental}. 

Conventional DCF are suitably constructed optical fibers with an appropriate refractive index profile such that they exhibit the desired dispersion value at the wavelength of operation. Since the dispersion coefficient of the transmission optical fiber is usually positive, the conventional DCF should exhibit negative dispersion coefficient. They should also have dispersion slope matching, low bend loss, low propagation loss, and relatively large mode effective area \cite{thyagarajan2007modeling, gruner2005dispersion}. Design trade-offs to meet some of these requirements are necessary, e.g., small dispersion coefficient is usually characterized by small mode effective area and consequently large nonlinear effects and vice versa \cite{thyagarajan2007modeling, gruner2005dispersion, ramachandran2007fiber}. In this letter, we take an opposite direction and show how an optical fiber with high positive dispersion coefficient can also be used to compensate for the dispersion. In fact, our compensating procedure is an adoption of an iterative method in signal processing that uses a given system to implements its inverse system \cite{marvasti2012nonuniform}. The proposed dispersion compensating structure is a cascade repetition of a sub-system and its compensation capability is improved by higher number of cascaded sub-systems in the structure. Furthermore, the capability of the proposed dispersion compensating technique is a trade-off between the transmission length and bandwidth. It is noteworthy that the introduced structure can simultaneously compensate for dispersion, dispersion slope, and other high order dispersions. Photonic Crystal Fibers (PCF) with high positive dispersion and low attenuation coefficients can be considered as a main candidate for constructing the offered compensating module \cite{bjarklev2012photonic, gerome2004design, russell2006photonic}. High positive dispersion and low attenuation coefficients of these PCFs improve the compensation ability of our structure by providing dispersion compensation with lower latency, attenuation, and sensitivity to nonlinear effects in comparison with the conventional DCFs. 

The rest of the paper is organized as follows. In Section \ref{Sec_II}, we present a brief survey of chromatic dispersion and its mathematical model. Following a concise review of the mentioned iterative method, we introduce and analyze our new dispersion compensating system in Section \ref{Sec_III}. In Section \ref{Sec_IV}, the performance of the proposed dispersion compensating structure is verified using simulation results. Finally, we conclude the paper in Section \ref{Sec_V}.

\section{Chromatic Dispersion}\label{Sec_II}
One of the most significant impairments in fiber optics communication systems is chromatic dispersion. The light pulses propagating along the optical fiber become distorted because different spectral components of the signal travel with different speed and hence experience different propagation delay during transmission. This means those parts of the signal will reach the receiver at different time instants which results in a temporal pulse distortion and broadening. 
Dispersion is characterized by the parameter $D$ which describes how the pulse is broadened. There are two physical issues accounting for the chromatic dispersion in  optical fibers. The first one is material dispersion which is due to the fact that the core and cladding are made of dispersive materials and this means that the refractive index is frequency dependent. The second one is waveguide dispersion which is caused by the frequency dependence of the propagation constant to the geometry and design of the optical fiber. Assuming that the fiber is a cylindrical dielectric waveguide along the z-axis, the
wave propagation in the frequency domain along the positive z-coordinate is given by \cite{ramachandran2007fiber}:
\begin{equation}
E(z,\omega) = E(0,\omega)e^{-j\beta(\omega)z}
\end{equation}
where $E(0,\omega)$ and $E(z,\omega)$ are pulse fields at frequency $\omega$ and distances $0$ and $z$ with respect to the origin, respectively and $\beta(\omega)$ is the fiber frequency-dependent propagation constant. Assuming that the spectral width $\Delta \omega=\omega-\omega_0$ is much smaller than the carrier frequency $\omega_0= \frac{2\pi c}{\lambda_0}$, the Fourier series expansion of the propagation constant around $\omega_0$ is:
\begin{equation}\label{beta}
\beta(\omega)=n(\omega)\frac{\omega}{c}=\beta_0 + \beta_1 (\Delta \omega)+ \frac{\beta_2}{2} (\Delta \omega)^2+\cdots
\end{equation}
where the series coefficients are written as $\beta_n = \frac{\partial^n\beta}{\partial \omega^n}\vert_{\omega = \omega_0}$. The group delay per unit of length is $\tau(\omega)=\frac{\partial\beta}{\partial \omega}$.
The first term of \eqref{beta} represents a frequency independent phase rotation that can be disregarded for the propagation of the pulse. The second parameter $\beta_1$ is equal to the group delay per unit of length at the carrier frequency and also equals to the inverse of the group velocity at the carrier frequency i.e. $\beta_1= \frac{1}{V_g(\omega_0)}, V_g(\omega) = \frac{c}{n(\omega)}$ ($n(\omega)$ is the frequency dependent refractive index). The third term of \eqref{beta} describes first-order chromatic dispersion and sometimes it is called the group-velocity dispersion. Chromatic dispersion parameter equals to $\beta_2 = \frac{d\tau}{d\omega}\vert_{\omega = \omega_0}$ and it is responsible for linear variation of group delay with frequency. Commonly, chromatic dispersion parameter is characterized in terms of $\Delta \lambda$ instead of $\Delta \omega$:
\begin{equation}
D= \frac{d\tau}{d\lambda}=-\frac{2\pi c}{\lambda_0^2}\beta_2
\end{equation}
where $D$ is the mentioned dispersion coefficient and its unit is usually expressed as $ps/nm/km$. Typical values of the main dispersion parameters in standard single mode fiber at the wavelength $\lambda_0 = 1550 nm$ are $\beta_2 = -21 ps^2/km$ and $D = 17 ps/nm/km$. As previously mentioned, the total dispersion parameter is given by the sum of both the material dispersion $D_m$ and the waveguide dispersion $D_{w}$ i.e. $D = D_m + D_{w}$. Other parameters $\beta_n, n=3,4,\cdots$ are related to high order dispersions and usually neglected in comparison with the main dispersion parameter $\beta_2$. Clearly, the effect of dispersion on pulse propagation through the optical fiber is totally characterized by the dispersion transfer function $H_D(\omega)$:
\begin{equation}
H_D(\omega) = \exp(-jz\sum\limits_{i=2}^{\infty}\frac{\beta_i}{i!}(\Delta\omega)^i)
\end{equation}
\section{Proposed Dispersion Compensating System}\label{Sec_III}
\begin{figure}[t!]
\centering
\includegraphics[scale = 0.6]{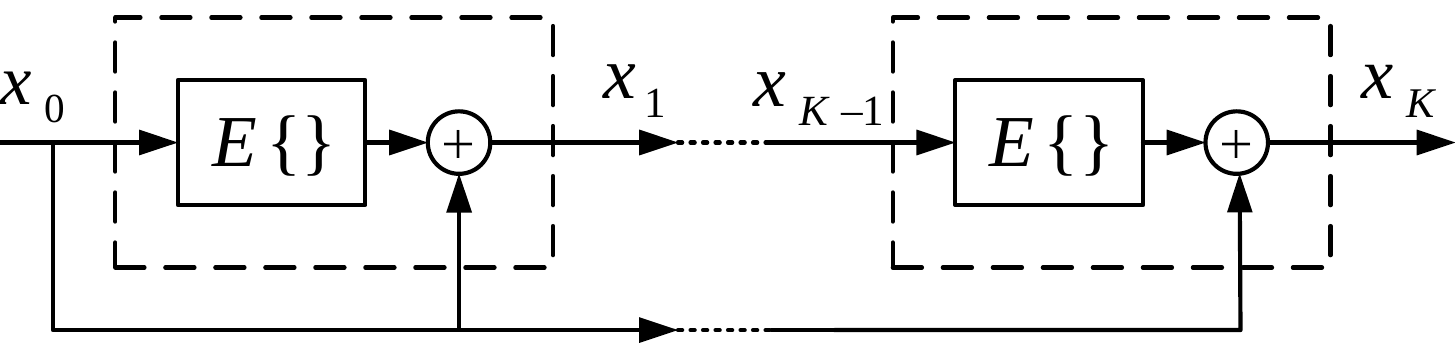}
\caption{Direct implementation of the iterative method which is a cascade repetition of the highlighted part. Each highlighted part is an error operator followed by an add operation.}\label{iter_method}
\vspace{-0.75cm}
\end{figure}

\begin{figure}[t!]
\centering
\includegraphics[scale = 0.6]{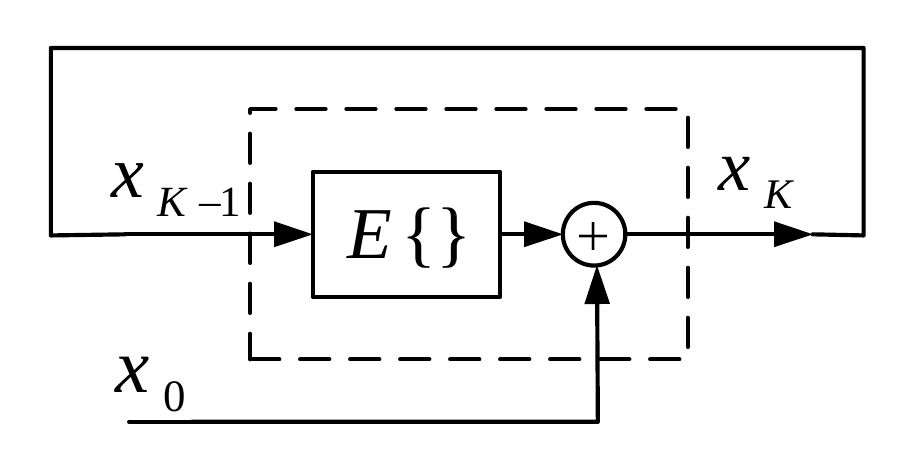}
\caption{Feedback implementation of the iterative method which is a closed loop system constructed of the highlighted part of Fig. \ref{iter_method}. Input signal should cycle $K$ times in the feedback loop to provide an output equivalent to the output of the direct implementation with $K$ repetition of the highlighted part.}\label{iter_method_feedback}
\vspace{-0.75cm}
\end{figure}

The iterative method is a systematic technique in which successive operations of a given operator is used to provide an estimate of its corresponding inverse operation \cite{marvasti2012nonuniform}. Consider an arbitrary operator named $H \lbrace \rbrace$ and let $I \lbrace \rbrace$ be the identity operator. Defining $E \lbrace \rbrace = I \lbrace \rbrace - H \lbrace \rbrace$ as error operator, $E^k \lbrace \rbrace$ means $k$ consecutive operations of the error operator. It can be shown that $H^{-1}\lbrace \rbrace$, the inverse of the operator $H\lbrace \rbrace$, can be calculated as follows:
\begin{equation}\label{iter_1}
I + E + E^2 + ... + E^k + ...  = H^{-1} 
\end{equation}
provided that $E \lbrace \rbrace$ is linear and its operator norm $\vert \vert E \vert \vert = \vert \vert I - H \vert \vert $ is less than $1$ \cite{marvasti2012nonuniform}. For sufficiently large values of $k$, \eqref{iter_1} provides an approximation of the inverse operator $H^{-1}\lbrace \rbrace$. Fig. \ref{iter_method} shows a systematic realization for \eqref{iter_1}. As shown in Fig. \ref{iter_method_feedback}, the inverse operator can also iteratively be realized by feedbacking the output of the highlighted part in Fig. \ref{iter_method} to its input. For speeding up or down the convergency or extending the convergence region, a scaling factor $\mu$ can be included in the definition of the error operator $E\lbrace \rbrace=I\lbrace \rbrace - \mu H \lbrace \rbrace$. Increasing $\mu$ speeds up the convergency at the cost of reducing the convergence region \cite{marvasti2012nonuniform}.
\begin{figure}[t!]
\centering
\includegraphics[scale = 0.6]{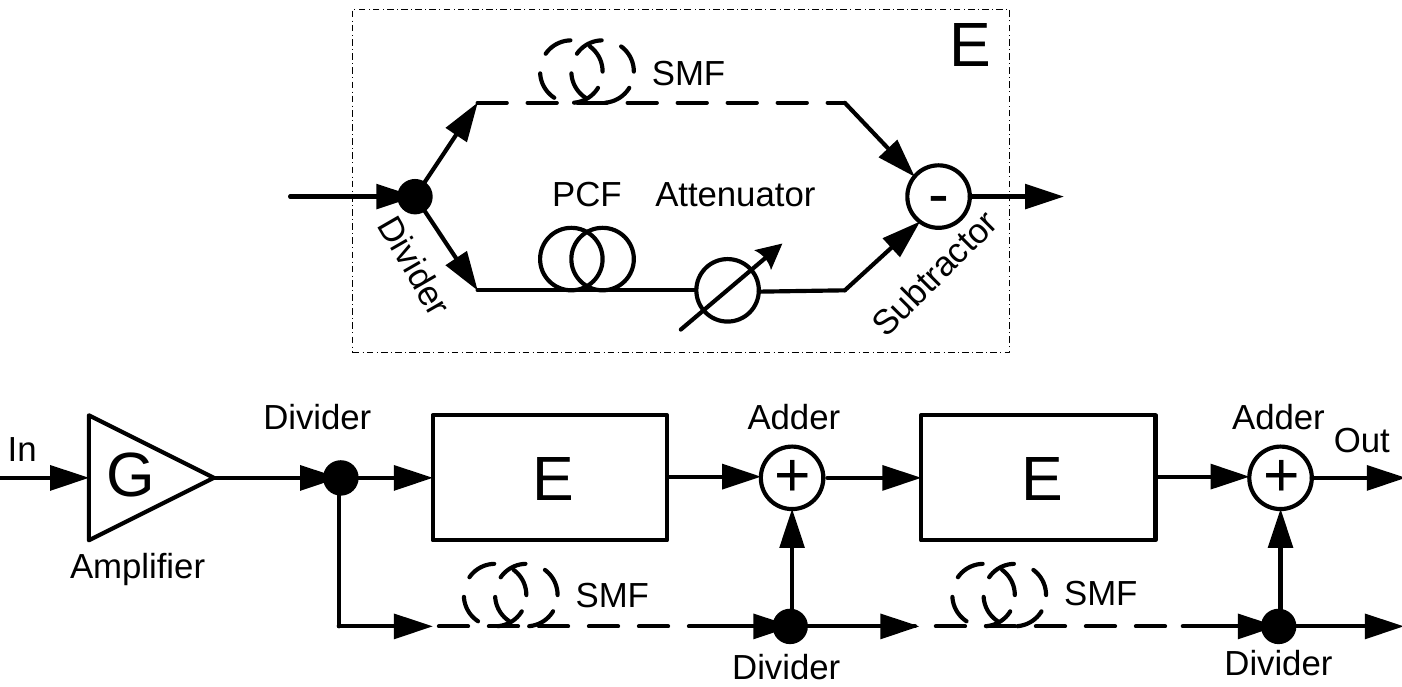}
\caption{Block diagram of the proposed dispersion compensating structure including $2$ cascaded sub-systems $E$.}\label{dcm}
\vspace{-0.75cm}
\end{figure}
Now, consider the structure shown in Fig. \ref{dcm}. This structure is an optical adoption of the introduced iterative method that implements the inverse of the chromatic dispersion transfer function using high-positive dispersive optical fibers. Sub-system $E(\omega)$ gets an input optical signal, divides it between two optical fibers and construct the output optical signal by subtracting the outputs of the optical fibers. We assume the optical fiber in up branch of the sub-system is an ordinary optical fiber with characterizing parameters $\beta_i^{SMF}, i=0,1,2,...$ and length $L$ while the optical fiber in down branch is a special optical fiber with characterizing parameters $\beta_i^{PCF}, i=0,1,2,\cdots$ and length $L$ such that $\beta_i^{PCF}\approx \beta_i^{SMF}, i = 0,1$ and $\frac{\beta_i^{PCF}}{\beta_i^{SMF}} \gg 1, i = 2,3,\cdots$. A well-designed high-positive dispersive PCF can satisfy the requirements of the optical fiber in the down brand of the sub-system. If the attenuator in the down branch has an attenuation coefficient $\alpha$, the transfer function of the sub-system is given by:
\begin{align}\label{edis}
\nonumber & E(\omega)  = \frac{1}{\sqrt{2}}e^{-jL\sum\limits_{i=0}^{\infty}\frac{\beta_i^{SMF}}{i!}(\Delta\omega)^i}-\frac{\sqrt{\alpha}}{\sqrt{2}}e^{-jL\sum\limits_{i=0}^{\infty}\frac{\beta_i^{PCF}}{i!}(\Delta\omega)^i}\\
& \nonumber \approx \frac{1}{\sqrt{2}}e^{-jL(\beta_0^{SMF}+\beta_1^{SMF}\Delta\omega)}(1-\sqrt{\alpha}e^{-jL\sum\limits_{i=2}^{\infty}\frac{\beta_i^{PCF}}{i!}(\Delta\omega)^i}) \\
& = \frac{1}{\sqrt{2}}e^{-jL(\beta_0^{SMF}+\beta_1^{SMF}\Delta\omega)}E_D(\omega)
\end{align}
In Fig. \ref{dcm}, the optical fiber below each embedded sub-system has the same characterizing parameters as the optical fiber in the up branch of the sub-system. Referring to \eqref{edis} and assuming $K$ cascaded copies of the analyzed sub-system, one can easily check that the total transfer function of the system $H^{-1}_{D}(\omega)$ is simplified as:
\begin{equation}
H^{-1}_{D}(\omega) \approx\frac{\sqrt{G}}{2^{\frac{K+1}{2}}}e^{-jKL(\beta_0^{SMF}+\beta_1^{SMF}\Delta\omega)} \sum\limits_{k=0}^K E_D^k(\omega)
\end{equation}
For sufficiently large number of cascaded sub-systems $K$ and for $\vert E(\omega) \vert < 1$:
\begin{equation}
H^{-1}_{D}(\omega) \approx \frac{\sqrt{G}}{\sqrt{\alpha}2^{\frac{K+1}{2}}}e^{-jKL(\beta_0^{SMF}+\beta_1^{SMF}\Delta\omega)} e^{jL\sum\limits_{i=2}^{\infty}\frac{\beta_i^{PCF}}{i!}(\Delta\omega)^i}
\end{equation}
which is the desired inverse transfer function of the chromatic dispersion with a causal delay coefficient. Assume we want to compensate for the chromatic dispersion of an optical fiber with characterizing parameters $\beta_i^{FIB}, i=0,1,2,\cdots$ and length $z$. If we set $\beta_i^{PCF}L=\beta_i^{FIB}z, i = 2,3,\cdots$ and $G=\alpha 2^{K+1}$, the proposed structure can totally remove the dispersion if:
\begin{equation}\label{conv_con}
\vert E_D(\omega) \vert = \vert 1-\sqrt{\alpha}e^{-jz\sum\limits_{i=2}^{\infty}\frac{\beta_i^{FIB}}{i!}(\Delta\omega)^i}) \vert< 1
\end{equation}
Neglecting high order dispersion coefficients $\beta_i^{FIB}, i=3,4,\cdots$  against the main dispersion coefficient $\beta_2^{FIB}$, \eqref{conv_con} implies that we have the following trade-off between transmission length and bandwidth for a stable dispersion compensation:
\begin{equation}\label{trad}
 \frac{\vert\beta_2^{FIB}\vert}{2}(\Delta \omega)^2 z  < cos^{-1}(\frac{\sqrt{\alpha}}{2})
\end{equation}
It is noteworthy that the optical amplifier of the proposed structure can be merged with the receiver amplifier and its other parts can totally be constructed using passive optical elements. Furthermore, the length of the required compensating fibers $L$, the compensation delay, and the compensation attenuation are reduced for higher values of  $\vert \beta_2^{PCF} \vert$.
\section{Simulation Results}\label{Sec_IV}
\begin{figure}[t!]
\centering
\includegraphics[scale = 0.6]{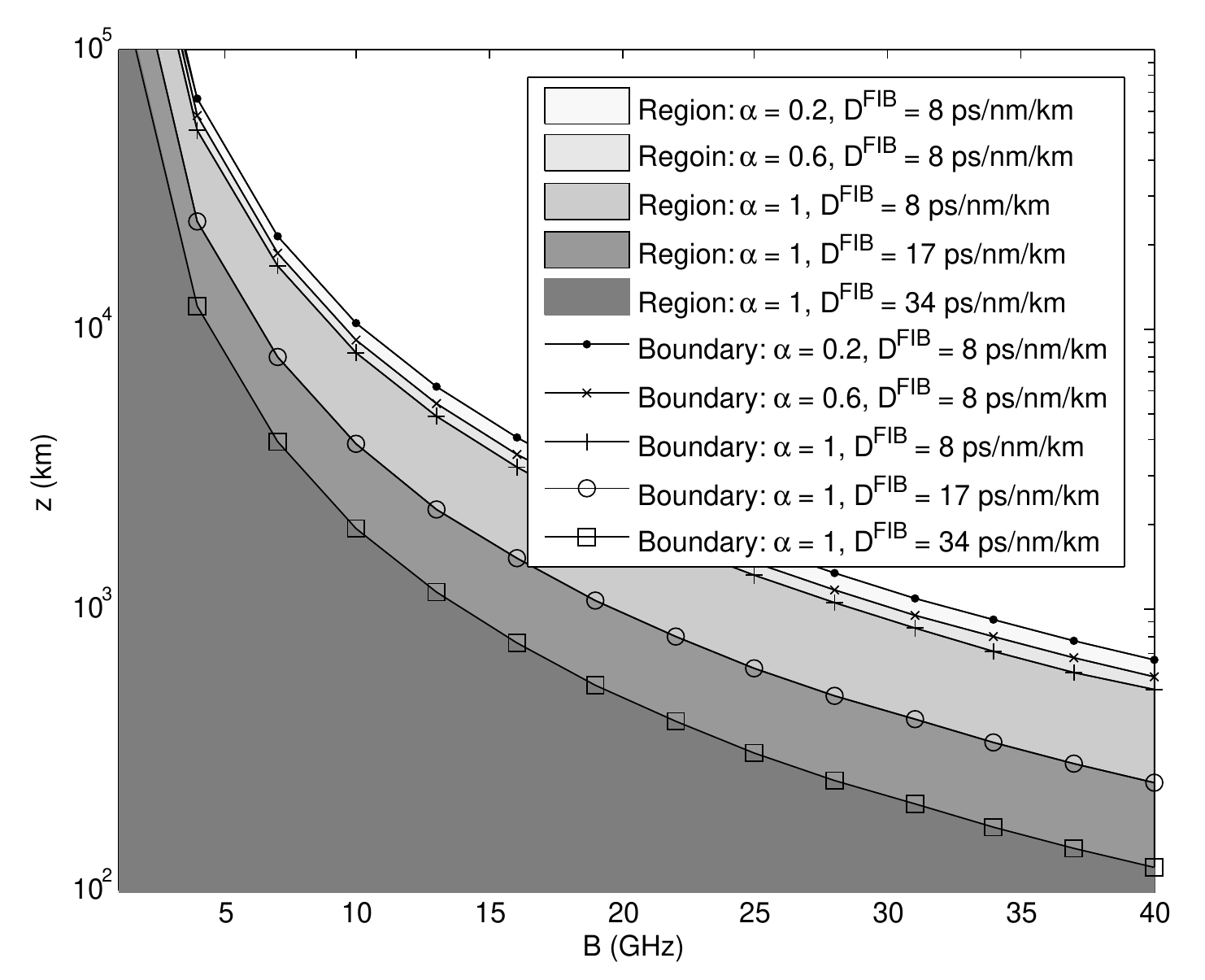}
\caption{2D region of transmission length $z$ and bandwidth $B$ values for which the proposed system can stably compensate for dispersion.}\label{tradeoff}
\vspace{-0.75cm}
\end{figure}

\begin{figure}[t!]
\centering
\includegraphics[scale = 0.6]{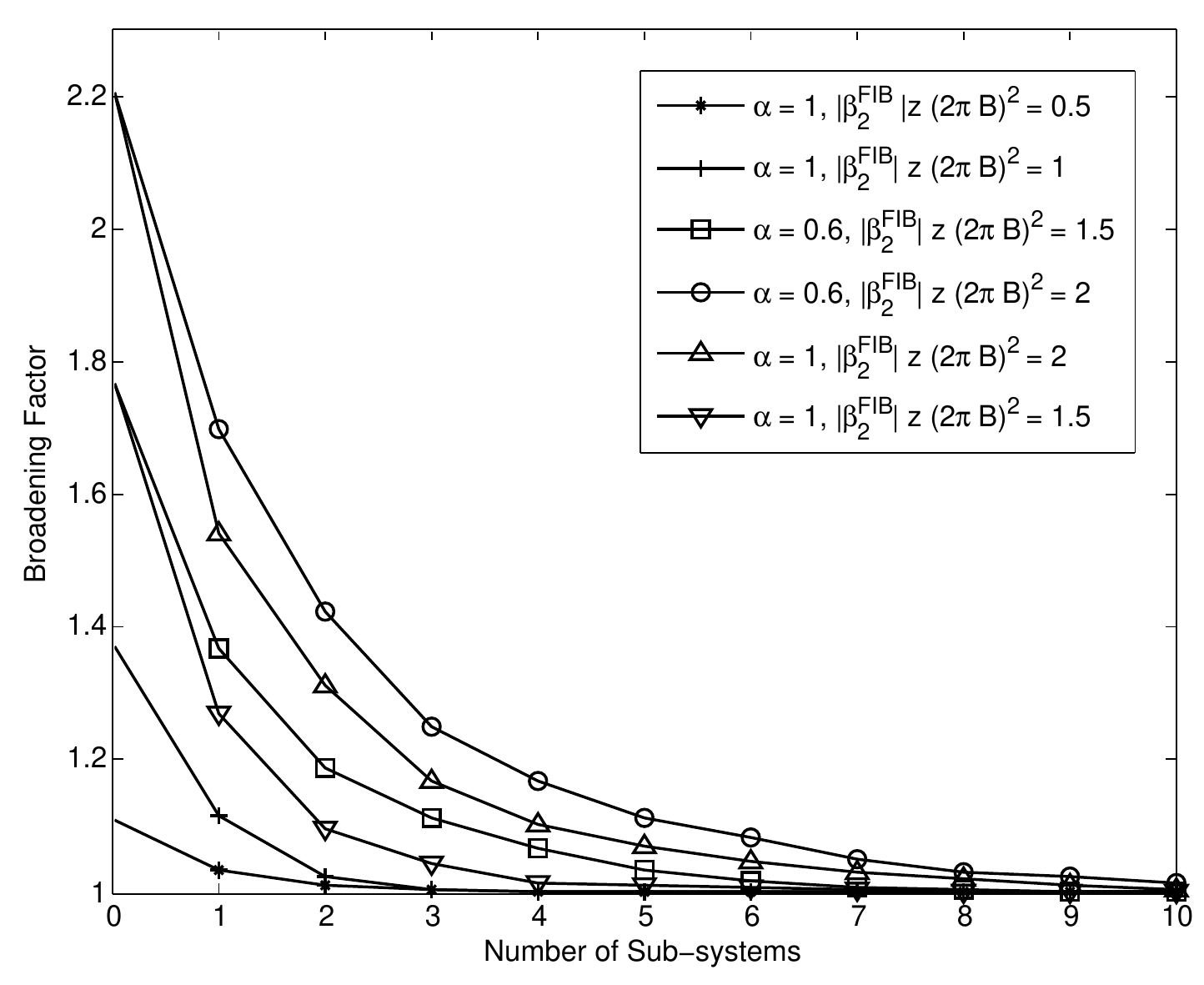}
\caption{Broadening factor for a sinc-shaped optical pulse conveyed by carrier wavelength $\lambda_0 = 1550 nm$ in terms of the number of cascaded sub-systems $E$ in the compensating structure for various values of $\vert \beta_2^{FIB}\vert z(2\pi B)^2$ and attenuation coefficient $\alpha$.}\label{iter}
\vspace{-0.75cm}
\end{figure}

Fig. \ref{tradeoff} shows the 2D region of transmission length $z$ and bandwidth $B=max\{\vert \Delta \omega \vert \}/ \pi$ values for which the introduced system can stably compensate for the dispersion. We refer to this region as convergence region. As illustrated, the boundary of this region is governed by the explicit trade-off between transmission length and bandwidth in \eqref{trad} and its area is extended for lower values of dispersion coefficient $D^{FIB}$ or attenuation coefficient $\alpha$. Now, consider a sinc-shaped optical pulse with a given zero-to-zero pulse width and its corresponding bandwidth $B$. This pulse is conveyed by carrier wavelength $\lambda_0 = 1550 nm$ over an optical fiber having the second order dispersion parameter $\beta_2^{FIB}$ and length $z$. We use the proposed system to compensate for the accumulated dispersion in sinc-shaped pulse propagated through this optical fiber. Fig. \ref{iter} shows the simulated broadening factor of the compensated pulse (i.e. the ratio of the received pulse width to the transmitted pulse width) in terms of the number of embedded sub-systems $E$ in the compensating system for various values of $\beta_2^{FIB}z(2\pi B)^2$ and attenuation coefficient $\alpha$. Clearly, the compensation performance is increased for higher number of cascaded sub-systems. Furthermore, a pulse with lower bandwidth and transmission distance needs lower number of sub-systems to get a desired dispersion compensated level. The simulation results also show decreasing the attenuation coefficient $\alpha$ decreases the convergence speed but as interpreted from Fig. \ref{tradeoff}, this can extend the convergence region to include a desired pair of transmission length and bandwidth. 


Assume that we desire to compensate for the accumulated dispersion of a single channel optical signal with $3 GHz$ bandwidth and $1550 nm$ carrier frequency that propagates through a $130 km$ standard single mode optical fiber such that a broadening factor of $1.1$ is achieved at the receiver side. Since $\vert \beta_2^{FIB}\vert z(2\pi B)^2 \approx 1$, Fig. \ref{iter} shows that our proposed dispersion compensating module including $1$ embedded sub-system can provide the desired compensated broadening factor. If $D^{PCF} \approx 2200 ps/nm/km$ (or equivalently $\beta_2^{PCF} \approx -2718 ps^2/km$), the signal propagation path of the proposed compensating structure will be around $1 Km$. As simulation results show, the favorite level of the compensated broadening factor can also be obtained using a sample DCF with $D^{DCF} = -250 ps/nm/km$ and approximated propagation path of $7 km$ \cite{gruner2005dispersion}. The propagation path in the proposed module is shorter than its counterpart DCF and hence it has a potential to provide lower attenuation and delay during the compensation process.

\section{Conclusion}\label{Sec_V}
In this paper, we took an opposite direction with respect to the conventional DCFs and showed an optical fiber with high positive dispersion coefficient can also be used for dispersion compensation. We proposed an optical structure based on a well-known iterative algorithm in signal processing in which dispersion inverse transfer function is implemented using high-positive dispersive optical fibers. We showed that the dispersion compensation capability of the proposed module is a trade-off between transmission length and bandwidth and is enhanced for the compensating structure including more dispersion compensating sub-systems. We also specified how system parameters should change to stabilize or speed-up the system performance. Generally, the concepts and ideas behind the proposed structure can be used in developing other innovative optical modules such as optical filters and other optical impairment compensating structures.


%





\ifCLASSOPTIONcaptionsoff
  \newpage
\fi



\bibliographystyle{IEEEtran}
\bibliography{References}
\end{document}